\documentclass[acmsmall, screen]{acmart}
\usepackage{tcolorbox}
\usepackage{multirow} 
\usepackage{colortbl}
\DeclareMathOperator*{\argmax}{arg\,max}

\AtBeginDocument{%
  \providecommand\BibTeX{{%
    \normalfont B\kern-0.5em{\scshape i\kern-0.25em b}\kern-0.8em\TeX}}}

\setcopyright{acmcopyright}
\copyrightyear{2023}
\acmYear{2023}
\acmDOI{XXXXXXX.XXXXXXX}

\newcommand{\smallsection}[1]{\noindent {\bf \underline{#1}}.\hspace{1mm}}
\newcommand{\ea}{\textit{et al.}}

\newcommand{\find}[1]{
\begin{tcolorbox}[leftrule=1mm,toprule=0mm,bottomrule=0mm,left=1pt,right=2pt,top=2pt,bottom=2pt]
\em #1
\end{tcolorbox}
}

\begin{document}

\title{On the Reliability and Explainability of Language Models for Program Generation}

\author{Yue Liu}
\email{yue.liu1@monash.edu}
\affiliation{%
  \institution{Monash University}
  \streetaddress{Wellington Road}
  \city{Clayton}
  \state{Victoria}
  \country{Australia}
}

\author{Chakkrit Tantithamthavorn}
\email{chakkrit@monash.edu}
\authornote{Corresponding Author}
\affiliation{%
  \institution{Monash University}
  \streetaddress{Wellington Road}
  \city{Clayton}
  \state{Victoria}
  \country{Australia}
}

\author{Yonghui Liu}
\email{Yonghui.Liu@monash.edu}
\affiliation{%
  \institution{Monash University}
  \streetaddress{Wellington Road}
  \city{Clayton}
  \state{Victoria}
  \country{Australia}
}

\author{Li Li}
\email{lilicoding@ieee.org}
\affiliation{%
  \institution{Beihang University}
  \streetaddress{Xueyuan Road}
  \city{Beijing}
  \country{China}
}

\renewcommand{\shortauthors}{Liu~\ea}

\begin{abstract}
Recent studies have adopted pre-trained language models, such as CodeT5 and CodeGPT, for automated program generation tasks like code generation, repair, and translation. 
Numerous language model-based approaches have been proposed and evaluated on various benchmark datasets, demonstrating promising performance. 
However, there is still uncertainty about the reliability of these models, particularly their realistic ability to consistently transform code sequences. 
This raises the question: are these techniques sufficiently trustworthy for automated program generation?
Consequently, Further research is needed to understand model logic and assess reliability and explainability. 
To bridge these research gaps, we conduct a thorough empirical study of eight popular language models on five representative datasets to determine the capabilities and limitations of automated program generation approaches.
We further employ advanced explainable AI approaches to highlight the tokens that significantly contribute to the code transformation.
We discover that state-of-the-art approaches suffer from inappropriate performance evaluation stemming from severe data duplication, causing over-optimistic results.
Our explainability analysis reveals that, in various experimental scenarios, language models can recognize code grammar and structural information, but they exhibit limited robustness to changes in input sequences. 
Overall, more rigorous evaluation approaches and benchmarks are critical to enhance the reliability and explainability of automated program generation moving forward. 
Our findings provide important guidelines for this goal.
\end{abstract}



\maketitle

\section{Introduction}
Program generation (e.g., code repair and code translation) involves producing new source code sequences to facilitate software development and maintenance. 
These activities require software engineers to manually analyze, understand, and even execute programs to ensure the quality of newly developed code.
Thus, automated learning-based program generation has been proposed to automatically generate new code based on context.
Recently, language models (LMs) have shown promise in semantic modeling and natural language understanding~\cite{vaswani2017attention,brown2020language}. 
Due to similarities between text and source code~\cite{chirkova2021empirical}, language models (e.g., LSTM and Transformer) have gained immense research interest for automated code generation and understanding~\cite{li2022closer, tufano2019learning}.
Specifically, pre-trained LMs for code (e.g., CodeT5~\cite{wang2021codet5}, and CodeBERT~\cite{feng2020codebert}), typically pre-trained on massive unlabeled code corpora using self-supervised objectives, can learn generic source code representations. 
These can then be transferred to diverse downstream tasks, decreasing pre-training costs and improving performance on code understanding and generation~\cite{wang2022bridging}.
Consequently, pre-trained LMs have become state-of-the-art for many code intelligence tasks, such as code review~\cite{li2022codereviewer, hong2022commentfinder}, code summarization~\cite{wan2022they}, code completion~\cite{wan2022they,liu2020multi}, and program repair~\cite{zhang2022coditt5}. 

Despite the effectiveness and advanced capabilities of LMs for code demonstrated in prior work, these powerful program generation models are not yet reliable enough for practical application~\cite{zeng2022extensive, she2023pitfalls}.
For example, prior studies~\cite{schuster2021you, yang2022natural, zeng2022extensive} have shown that LMs are extremely vulnerable to adversarial attacks, where attackers can induce incorrect outputs through minor perturbations.
Additional analysis and demonstration of model robustness are clearly needed. 
Furthermore, impressive results in lab-only settings may not translate to real-world efficacy~\cite{she2023pitfalls}. 
We discuss three significant issues and implications that must be addressed before widespread adoption of program generation models:

\noindent\textbf{Insufficient interpretability analysis.}
One severe drawback of prior research is a lack of explainability analysis on language models for code.
In contrast to linear models, LMs have complex architectures (e.g., standard CodeT5 with 12 layers, 220M parameters), making it challenging to explain specific predictions~\cite{tantithamthavorn21actionable, tantithamthavorn21xai, 10109341, 10109328}.
In other words, the recommendations made by LMs are opaque to practitioners and researchers, who are unlikely to accept code recommendations without evidence, especially for non-trivial cases ~\cite{rabin2021understanding, jiarpakdee2021practitioner, mohammadkhani2023systematic}.
For example, while prior studies have proven that LMs can automatically update code to fix bugs, 
which input characteristics contribute to the prediction is largely overlooked~\cite{zhang2022coditt5, zeng2022extensive}.
This gap in interpretability analysis significantly hinders the real-world adoption of program generation models.

\noindent\textbf{Experimental bias.} 
While prior research has demonstrated the effectiveness of LMs on code-based tasks, their performance evaluations can be susceptible to experimental bias~\cite{zeng2022extensive}. 
A common issue is dataset bias~\cite{chakraborty2021multi, sun2022importance, zhang2023slice}, where most code snippets are collected from open-source projects in a compromised manner that potentially introduces noise.
For example,  Zhang~\ea~\cite{zhang2023slice} trained Transformer-based methods on a noisy dataset containing both bot-generated and trivial code commit messages, achieving a 42.4\% BLEU-4 score. However, removing the noisy data resulted in a sharp performance drop to 26.2\% BLEU-4.
Therefore, the actual capabilities of LMs trained on such biased datasets are difficult to accurately evaluate. They likely face severe performance degradation when applied in practice versus lab settings.

\noindent\textbf{Poor practicability.}
While substantial research indicates promise for using LMs in automated program generation, most evaluation results demonstrate these techniques are not yet practical for real-world application. 
For example, Tufano~\ea~\cite{tufano2019learning} proposed a Transformer-based automated code change approach, achieving just 21.16\% accuracy. 
With an accuracy of 21\% (i.e., successfully predicted code transformations), there are too many false negatives, which means practitioners can never know whether the code update fits or not. 
Further experiments showed increasing the beam search size from one to ten brings a significant performance improvement (i.e., 21.16\% to 36.02\%).
However, picking the best sequence among ten candidates remains challenging for practitioners.

To the best of our knowledge, no systematic research study has investigated the reliability and explainability of language models for program generation.
In response to the observations and concerns raised above, we conduct the first comprehensive empirical evaluation of popular pre-trained language models on automated program generation tasks. 
Specifically, we adopt eight mainstream pre-trained models, i.e., T5~\cite{raffel2020exploring}, CodeT5~\cite{wang2021codet5}, CoTexT~\cite{phan2021cotext}, CodeTrans~\cite{elnaggar2021codetrans}, CodeGPT~\cite{lu2021codexglue}, CodeBERT~\cite{feng2020codebert}, CodeT5+~\cite{wang2023codet5+}, CodeReviewer~\cite{li2022codereviewer}.
We evaluated performance on four popular program generation tasks: code repair, code review, code translation, and text-to-code generation using five state-of-the-art benchmark datasets (i.e., \textit{Tufano~\ea~\cite{tufano2019learning}}, \textit{Bugs2Fix~\cite{tufano2019empirical}}, \textit{CodeReview~\cite{li2022codereviewer}}, \textit{CodeTrans-Dataset~\cite{lu2021codexglue}}, \textit{CONCODE~\cite{iyer2018mapping}}).
Our results confirm the superior accuracy of studied models reported in prior work, with some achieving nearly 70\% on certain datasets.
However, we discovered performance is skewed by inappropriate experimental design, yielding unreliable and unrealistic evaluations.
Specifically, duplication in benchmark datasets skews results. Some have explicit duplicates between training and testing sets, directly inflating performance. In addition, even where testing sets lack identical training examples, they may contain duplicated testing cases. This overlapping test data further distorts evaluation.
For code-to-code tasks, a concerning observation is that a substantial percentage of generated code exactly matches inputs, instead of generating refined or updated code sequences.
Moreover, explanation results suggest that program generation models can recognize code grammar and structural information.
However, they present poor robustness even to minor changes in input sequences.
In contrast to previous research, our findings indicate that automated program generation remains imperfect, with ample room for improvement through future work.

\noindent\textbf{\underline{Contribution.}} The main contributions of this paper are summarized as follows:
\begin{itemize}
    \item We conduct the first comprehensive benchmark study of pre-trained language models for program generation, evaluating reliability and explainability.
    \item Our analysis reveals significant experimental biases in prior work, including dataset duplication and overlapping inputs that inflate performance claims. Explanation analysis demonstrates models overlook critical tokens and lack robustness, highlighting key challenges for practical deployment.
    \item Results provide insights to guide future research toward more rigorous and reliable language models for neural program generation.
\end{itemize}

\noindent\textbf{\underline{Open Science.}} To support the open science initiative, we publish the studied dataset and a replication package, which are publicly available on GitHub\footnote{https://github.com/yueyueL/ProgramGen-LMs-Reliability}.

\noindent\textbf{\underline{Paper Organization.}}
The remainder of this paper is structured as follows: Section~\ref{sec:background} presents the background. Section~\ref{sec:exp_disign} details our experimental design. Section~\ref{sec:result} provides our experimental results and analysis. Section~\ref{sec:discussion} discusses the study's implications and potential threats to its validity. Related work is briefly introduced in Section~\ref{sec:related_work}, followed by a conclusion in Section~\ref{sec:conlcusion}.

\section{Background}
\label{sec:background}

\subsection{Automated Program Generation}
Automated program generation exploits language models to analyze and emulate the distribution patterns of both source code and natural text, enabling the generation of new code snippets.
As a result of the similarity of the data representation, program generation models inherit many techniques from natural language processing.
As a result, a variety of techniques from natural language generation, especially those involving Transformer-based models, have found increasing application in automated code-based tasks~\cite{hou2023large, fan2023large}.
In addition, language models like T5~\cite{raffel2020exploring} have demonstrated the ability to effectively learn code representation from unlabeled data to conduct a wide range of downstream tasks given supervised discriminative fine-tuning on specific tasks (e.g., code completion~\cite{takerngsaksiri2023syntax, wan2022they,liu2020multi}, code search~\cite{gu2018deep, sun2022importance}, code summarization~\cite{wan2022they}, code review~\cite{tufano2019learning, Patanamon2022AutoTransform, li2022codereviewer, Pornprasit23,thongtanunam2022autotransform}, API recommendation~\cite{chen2021holistic}, and vulnerability analysis~\cite{fu2022LineVul, fu2022vulrepair,fu2023aibughunter, fu2023vision, fu2023vulexplainer, fu2023chatgpt}). 

In this work, our focus is exclusively on automated program generation tasks that involve the transformation of a source code snippet or natural language into a new or modified program snippet, since their strengths and limitations have not been thoroughly studied. 
We describe program generation models for the task of mapping a source sequence $X = [x_{1}, ..., x_{M}]$ (where $M$ represents the length of the source sequence) to a target sequence $Y = [y_{1}, ..., y_{N}]$ (where $N$ signifies the target sequence length).
Specifically, the raw source sequences will be first preprocessed and mapped into a sequence of embeddings $(x_{1},...,x_{M})$.
Next, the Transformer-based models jointly train the encoder and decoder for comprehensive source code modeling.
In particular, program generation models are usually composed of Encoder layers and Decoder layers, where the Encoder takes a sequence of code tokens as input in order to map an initial method $X = [x_{1}, ..., x_{M}]$ into a fixed-length intermediate hidden state $H = [h_{1}, ..., h_{M}]$.
Then, the decoder takes the hidden state vector $H$ as an input to generate the output sequence of tokens $Y = [y_{1}, ..., y_{N}]$. 
We note that $M$ (i.e., the length of the input sequence) and $N$ (i.e., the length of the output sequence) can be different.
To optimize the mapping, the parameters of the model are updated using the training dataset with the following equation to maximize the conditional probability:

\begin{equation*}
p(Y \mid X) = p(y_{1}, ..., y_{m} \mid x_{1}, ..., x_{n})
                    = \prod_{i=1}^{m} p(y_{i} \mid H, y_{1}, ..., y_{i-1})
\end{equation*}

Many pre-trained language models (e.g., T5~\cite{raffel2020exploring}, CodeT5~\cite{wang2021codet5}, and CoTexT~\cite{phan2021cotext}) have been proposed and studied on various benchmarks and datasets.
These pre-trained models decrease the cost of pre-training a large-scale language model from scratch and improve the performance of various source code understanding and generation tasks~\cite{wang2022bridging}.
In other words, pre-trained code language models can be fine-tuned on different datasets for specific tasks, making them a state-of-the-art architecture in automated program generation. 
 
\subsection{Model Explanation}
\label{sec:model_explain_AI_approach}
Interpreting automated program generation models is critical for practitioners and researchers to comprehend and refine the decision-making processes inherent to these models. The application of explainable AI approaches serves to illustrate the learned behaviors of these models.

\noindent\textbf{Explainable Program Generation}
Consider an automated program generation model \( f \) that transforms a source sequence \( X=(x_{1},\ldots,x_{M}) \) into a predicted code sequence \( Y=(y_{1},\ldots,y_{N}) \), where \( M \) and \( N \) represent the lengths of the input and output sequences, respectively. The goal of an explainable AI approach is to ascertain the relevance \( r_{i} \)  of each token within the context of both the input and the sequence of tokens generated thus far.
This relevance can be formalized as \( r_{i}=(a^{1},\ldots,a^{M}, b^{1},\ldots,b^{i-1}) \), where \( a^{k} \) represents the importance of the \( k \)-th input token and \( b^{j} \) signifies the influence of the \( j \)-th token in the existing output sequence upon the generation of the next token \( y_{i} \). 
For each token \( y_{i} \) in the predicted sequence \( Y \), its feature importance is reflected not just by the input tokens but also by the contribution of all preceding tokens in the output sequence.

\begin{figure}[t]
  \centering
  \includegraphics[width=0.45\textwidth]{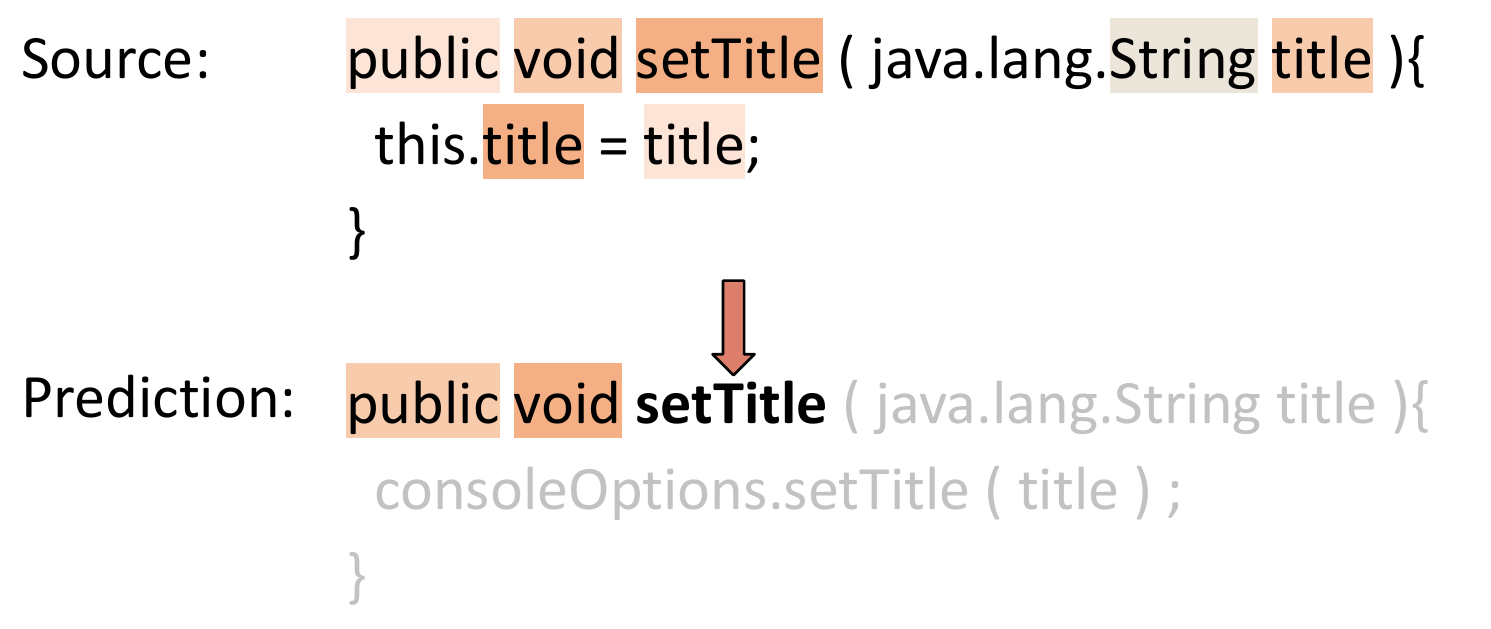}
  \caption{An explanation example for code review tasks, illustrating the generation of the "setTitle" token from a given source code sequence.}
  \label{fig:example_exp}
\end{figure}

Figure~\ref{fig:example_exp} presents an example of explainable AI applied to program generation. It displays the model's reasoning when generating the "setTitle" token, taking into account both the original input and the previously generated tokens. As depicted, tokens such as "setTitle" and "title" from the source sequence, and "void" from the existing generated tokens, are deemed highly important in this context. This visualization reinforces the sequence-to-sequence nature of program generation, where each token is generated in consideration of both the preceding and subsequent context.
In this section, we employ interpretable AI approaches that enable us to quantify the significance of each token conversion from the source code to the generated program sequence. By doing so, we aim to unpack the decision-making processes of program generation models, thereby offering insights into their predictive behavior.

\subsubsection{Attention-based analysis}
Most state-of-the-art pre-trained language models are based on the Transformer architecture~\cite{vaswani2017attention}, with the self-attention mechanism.
The attention mechanism, which provides a distribution of scores over the input tokens, has often been presented as showing the relative importance of the inputs. 
Specifically, the higher the attention weights, the more attention that is paid by the model.
Therefore, there have been many prior studies that employ the attention weights of pre-trained programming language models to explain model predictions~\cite{zhang2022diet, wan2022they, wang2021wheacha}.
Prior studies calculate the feature importance of each token by averaging the attention weights of all layers and heads.
However, variations in attention across different heads and layers, as noted by Wan~\ea~\cite{wan2022they}, suggest that each attention head focuses differently on various aspects of the source code, raising questions about the efficacy and accuracy of attention weights as explanatory tools.
Given the complexity and layered nature of pre-trained program generation models, and the ambiguity in determining the saliency of different attention weights for model predictions, we employ a model-agnostic interpretable approach in this study.

\subsubsection{SHAP}
SHAP (i.e., SHapley Additive exPlanations)~\cite{lundberg2017unified} is a popular black-box model-agnostic interpretable approach.
SHAP treats the candidate model $f$ as a black-box.
SHAP utilizes Shapley Values, a game theory-based approach, to approximate the relationship between the input and the output prediction.
Contrary to attention analysis, SHAP provides one feature importance vector to explain the whole relationship between source and target sequences. 
Consequently, we use the SHAP approach to analyze and understand language models in this study.
SHAP provides multiple explainable methods to generate SHAP values as an explanation.
In this study, we apply GradientExplainer to LM-based program generation models.
GradientExplainer is an extension of the integrated gradients approach and approximates SHAP values by computing the expectations of gradients by randomly sampling from the distribution of baseline/references.

\section{Experimental Design}
\label{sec:exp_disign}
In this section, we describe the setup and methodologies of our empirical study, focusing on the reliability and explainability of LM-based program generation.

\subsection{Language Models}
\label{sec:pretrained-models}
Recently, pre-trained language models for automated code understanding and generation tasks have been extensively studied in academia and industry.
These existing approaches can be categorized into three types: encoder-based models such as CodeBERT~\cite{feng2020codebert}, decoder-based models such as CodeGPT~\cite{lu2021codexglue}, and encoder-decoder-based models like CodeT5~\cite{wang2021codet5}. 
Although prior works~\cite{chakraborty2022natgen, li2022codereviewer, zeng2022extensive} have proven that encoder-based models and decoder-based models are not good at generation tasks, we include them in our study for completeness.
While there can be many potential state-of-the-art models to be studied, we chose eight representative pre-trained language models (i.e., T5~\cite{raffel2020exploring}, CoTexT~\cite{phan2021cotext}, CodeTrans~\cite{elnaggar2021codetrans}, CodeBERT~\cite{feng2020codebert}, CodeGPT~\cite{lu2021codexglue}, CodeT5~\cite{wang2021codet5}, CodeReviewer~\cite{li2022codereviewer}, and CodeT5+~\cite{wang2023codet5+}) that are publicly available for evaluation.

\noindent\textbf{T5~\cite{raffel2020exploring}.}
Raffel~\ea~\cite{raffel2020exploring} proposed the T5 (Text-To-Text Transfer Transformer) architecture, pre-trained on a large natural language corpus.
T5 has demonstrated state-of-the-art performance when fine-tuned for many NLP tasks. As T5 is only pre-trained on natural language data, we utilize it as a baseline to compare against pre-trained models for programming language.

\noindent\textbf{CoTexT~\cite{phan2021cotext}.}
utilizes the same architecture as T5 and is pre-trained on the CodeSearchNet Corpus~\cite{husain2019codesearchnet} and Google BigQuery~\cite{BigQuery}. CoTexT has achieved state-of-the-art results on code generation benchmarks~\cite{phan2021cotext}.

\noindent\textbf{CodeTrans~\cite{elnaggar2021codetrans}.} 
CodeTrans is a large pre-trained transformer-based encoder-decoder inspired by T5. 
Since the official online repository provides many versions, we use the version with the most downloads.
The CodeTrans model we used in this study is trained on 9,714 Java open-source projects from Github. 

\noindent\textbf{CodeBERT~\cite{feng2020codebert}.}
CodeBERT is an encoder-based model pre-trained on natural language and programming language corpora using RoBERTa architecture. 
CodeBERT is pre-trained on CodeSearchNet Corpus~\cite{husain2019codesearchnet}.
It has demonstrated strong performance on code search and summarization~\cite{zeng2022extensive}.

\noindent\textbf{CodeGPT~\cite{wang2021codet5}.}
CodeGPT is a decoder-only transformer model pre-trained on multiple programming languages.
CodeGPT is designed for code generation tasks like method name prediction, code completion, and code translation~\cite{lu2021codexglue}.

\noindent\textbf{CodeT5~\cite{wang2021codet5}.}
CodeT5 is a state-of-the-art unified pre-trained encoder-decoder programming language model. 
Inspired by T5, Wang~\ea~\cite{wang2021codet5} pre-trained the T5 architecture on eight programming languages together with their comments collected from open-source repositories.
CodeT5 has demonstrated promising performance on code-related generation tasks~\cite{hong2022commentfinder, le2022coderl, fu2022vulrepair}.

\noindent\textbf{CodeReviewer~\cite{li2022codereviewer}.} CodeReviewer is built on CodeT5 and is pre-trained on a large dataset of code updates and corresponding comments in code review scenarios.
CodeReviewer has proven prominent performance on code refinement by recent works~\cite{li2022codereviewer}.

\noindent\textbf{CodeT5+~\cite{wang2023codet5+}.}
CodeT5+ is an enhanced version of CodeT5, introducing architectural enhancements and advanced pretraining techniques like span denoising and contrastive learning. These innovations enable CodeT5+ models to achieve state-of-the-art performance across diverse code intelligence tasks, even zero-shot text-to-code generation\cite{wang2023codet5+}. 

\begin{table}[t]
\centering
\caption{Summary of Datasets Used for Different Tasks}
  \resizebox{0.85\linewidth}{!}{
\begin{tabular}{llllr}
\hline
\textbf{Task}          & \textbf{Subsets}          & \textbf{Category}         & \textbf{Language}                         & \textbf{Dataset Size} \\ \hline
 \begin{tabular}[c]{@{}l@{}}\textbf{\textit{Android\_S}}, \textbf{\textit{Android\_M}}, \\ \textbf{\textit{Google\_S}}, \textbf{\textit{Google\_M}}, \\ \textbf{\textit{Ovirt\_S}}, \textbf{\textit{Ovirt\_M}}\end{tabular} &Code Review            & Code-Code            & Java                                       & 21,774       \\ 
 \textbf{\textit{CodeReview}}                                                   &Code Review            & Code+Comment-Code    & \begin{tabular}[c]{@{}l@{}}Java, Python, Go, C++, C, \\ C\#, JavaScript, Php, Ruby\end{tabular} & 1.3M         \\ 
\textbf{\textit{B2F\_S}}, \textbf{\textit{B2F\_M}}                            &Code Repair            &  Code-Code            & Java                                       & 123,805      \\ 
 \textbf{\textit{Java2C\#}}, \textbf{\textit{C\#2Java}}                         &Code Translation       & Code-Code            & Java, C\#                                  & 11,500       \\ 
\textbf{\textit{CONCODE}}                                                       &Code Generation        & Text-Code            & Java                                       & 104,000      \\ \hline
\end{tabular}
}
\label{tab:dataset_summary}
\end{table}

\subsection{Downstream Tasks and Corresponding Datasets}
\label{sec:dataset}
Program generation tasks involve producing or generating sequences of tokens in a programming language. 
To better understand why language models perform program generation effectively, we extensively evaluate four different scenarios, in five datasets, for automated program generation as shown in Table~\ref{tab:dataset_summary}. 
The details of each scenario are described below.

\noindent\textbf{Code Review.} 
The objective of code review is to automatically implement code changes by developers during pull requests, including bug fixing, refactoring, and optimization.
Code review is a critical part of software development and serves as a common program generation task, extensively examined in prior research~\cite{li2022codereviewer, Patanamon2022AutoTransform, tufano2019learning}.
In our empirical study, we assess two different datasets related to code review tasks.

\begin{itemize}
    \item \textit{\textbf{Tufano~\ea~\cite{tufano2019learning}.}} Tufano~\ea~\cite{tufano2019learning} collected code review data from three large Gerrit~\cite{gerrit2022} code review repositories (i.e., \textit{Android}, \textit{Google}, and \textit{Ovirt}).
    This collection forms a code-to-code corpus, where the source code prior to the pull request is transformed into the target code, reflecting the changes post-review.
    The dataset is segregated according to method length into 10,783 \textit{Small} instances (with fewer than 50 tokens) and 10,991 \textit{Medium} instances (with 50 to 100 tokens). Consequently, six subsets are obtained: \textit{Android\_S}, \textit{Android\_M}, \textit{Google\_S}, \textit{Google\_M}, \textit{Ovirt\_S}, and \textit{Ovirt\_M}.

    \item \textit{\textbf{CodeReview~\cite{li2022codereviewer}.}}
    Li~\ea~\cite{li2022codereviewer} collected pull request data from the top popular open-source projects on GitHub.
    This expansive dataset mines the projects in nine programming language data, resulting in a collection of approximately 1.3 million pairs of pre- and post-review code.
    The dataset forms a  "code+text to code" corpus, where the input is the initial code along with associated commits, and the output is the revised code following the pull request.
\end{itemize}

\noindent\textbf{Code Repair.} 
Code repair aims to fix bugs in the code automatically.
This is one of the most common program generation tasks and has been widely investigated by prior work~\cite{tufano2019empirical,lu2021codexglue}. 

\begin{itemize}
    \item \textit{\textbf{Bugs2Fix.}} Tufano~\ea~\cite{tufano2019empirical} systematically extracted bug-fixing commit data from a large number of GitHub repositories, obtaining method-level pairs of buggy and fixed Java code snippets. This dataset presents a code-to-code transformation task, where the source is the buggy code and the target is the fixed code. The dataset is split into two subsets based on the code length, \textit{B2F\_S} for small fixes ($\leq 50$ tokens) and \textit{B2F\_M} for medium fixes ($> 50$ and $\leq 100$ tokens). Respectively, these subsets encompass 58,350 small and 65,455 medium bug-fix instances.
\end{itemize}

\noindent\textbf{Code Translation~\cite{nguyen2015divide}.} This task aims to translate code from one programming language to another while preserving its functionality. 

\begin{itemize}
    \item \textit{\textbf{CodeTrans-Dataset~\cite{lu2021codexglue}.}}
    This dataset, part of the CodeXGLUE benchmark~\cite{lu2021codexglue}, provides paired examples for code translation between Java and C\#, culminating in a total of 11,800 functionally equivalent method pairs. 
    This dataset enables the evaluation of translation from Java to C\# and vice versa, resulting in two distinct datasets: \textit{Java2C\#} and \textit{C\#2Java}.
\end{itemize}

\noindent\textbf{Code Generation.} Code generation refers to the creation of executable code based on natural language descriptions.  
\begin{itemize}
    \item \textit{\textbf{CONCODE~\cite{iyer2018mapping}.}}
     This dataset is widely used in code generation research, and collects examples from approximately 33,000 Java projects on GitHub. It encompasses 100,000 examples for training along with 4,000 examples for validation and testing. Each example is composed of a triple: a natural language description, code environments, and code snippets. This dataset exemplifies the "text-to-code" challenge, emphasizing the need for algorithms capable of understanding textual descriptions and transforming them into syntactically and semantically correct code.
\end{itemize}

In summary, our empirical study spans multiple task scenarios including code review, code repair, code translation, and code generation. 
We have obtained 12 distinct subsets for this purpose: \textit{Android\_S}, \textit{Android\_M}, \textit{Google\_S}, \textit{Google\_M}, \textit{Ovirt\_S}, \textit{Ovirt\_M}, \textit{CodeReview}, \textit{B2F\_S}, \textit{B2F\_M}, \textit{Java2C\#}, \textit{C\#2Java}, and \textit{CONCODE}.
These datasets facilitate the exploration of three input-output paradigms: code-to-code, code+text-to-code, and text-to-code, each pivotal to the domain of program generation.

\subsection{Evaluation Metrics}
To evaluate model performance, we primarily use accuracy (i.e. exact match rate or perfect prediction rate), which is the commonly-used metric for program generation tasks~\cite{Patanamon2022AutoTransform, li2022codereviewer, tufano2019empirical,tufano2019learning}.
Specifically, accuracy is calculated by dividing the number of perfectly predicted code snippets by the total snippets in the test set.
Additionally, we incorporate the BLEU-4 score from the NLP domain~\cite{papineni2002bleu} to evaluate the similarity between the generated and target code.
BLEU helps evaluate partial matches and overall fluency, complementing the exact match rate.
To further analyze potential data duplication within datasets, we introduce a modified BLEU-4 metric to quantify dataset-level similarity. This enables identifying and measuring repetitive patterns that could skew model training and evaluation.
For explanation results, we analyze and compare the average feature importance vectors $\bar{r}$. This provides insight into how models utilize input features for prediction.

\subsection{Implementation}
We build language models on top of two Python libraries, i.e., PyTorch~\cite{paszke2019pytorch} and Transformers~\cite{wolf2019huggingface}.
The studied models from Section~\ref{sec:pretrained-models} are obtained via the Transformers API.
To enable fair comparison, we use the 12-layer base version for all models, consistent with common practice in relevant literature~\cite{li2022codereviewer,zeng2022extensive}.
In this study, encoder-based models (CodeBERT) are appended with a transformer decoder to generate code regressively. Decoder (CodeGPT) and encoder-decoder-based models (CodeT5) directly generate code aggressively, as adopted by prior studies~\cite{lu2021codexglue,zeng2022extensive}.
For each dataset, we use identical training, validation, and test splits as well as fine-tuning approaches described in CodeXGLUE~\cite{lu2021codexglue} and original papers. 
The \textit{Bugs2Fix}, \textit{CodeTrans-Dataset}, and \textit{CONCODE} datasets were collected by CodeXGLUE, a benchmark for program understanding and generation. The \textit{Tufano~\ea} and \textit{CodeReview} datasets are in original format.
Regarding our explorations into explainable AI, we adopt the implementations provided by the Captum~\cite{kokhlikyan2020captum} and Ecco~\cite{alammar-2021-ecco} packages, both of which are designed for Python. 
Experiments are run on a machine with an AMD Ryzen 9 5950X 16-Core @ 3.4GHz CPU, 64GB RAM, and an NVIDIA GeForce RTX 3090 GPU with 24GB memory.

\subsection{Research Questions}
We structure our study around three key research questions to comprehensively evaluate language models for program generation:

\noindent\textbf{RQ1. How do language models perform on program generation tasks?}
Numerous pre-trained language models have been proposed, with some demonstrating promising capabilities on certain tasks. 
However, a systematic and extensive exploration of performance across diverse datasets and models is lacking. 
We undertake this broad investigation, evaluating the latest state-of-the-art models on tasks spanning code repair, review, translation, and generation.
 This RQ aims to undertake a comprehensive evaluation of LMs, examining their ability to not only replicate previous success but also generalize across distinct program generation tasks and datasets.

\noindent\textbf{RQ2. How reliable are automated program generation approaches?}
This research question critically analyzes the evaluation approaches used to assess automated program generation models, in order to identify potential experimental flaws or biases that could undermine performance evaluation.
Specifically, we investigate the representativeness and diversity of training and testing datasets. 
Performance heavily relies on dataset quality (i.e. "garbage in, garbage out")~\cite{she2023pitfalls}. 
Widespread duplication or lack of diversity could skew results.
In this RQ, we aim to uncover potential limitations that lead to overestimating or underestimating realistic capabilities. 
Addressing these concerns is vital for establishing robust and reliable benchmarking practices that will contribute to accurate characterization and continued improvement of automated program generation approaches.

\noindent\textbf{RQ3. Can we explain why automated program generation approaches can (or fail to) generate code sequences reliably?}
Only analyzing the generated sequences still does not determine why language models perform effectively or ineffectively.
The main reason is that what these pre-trained language models are based on to predict new code sequences is largely unknown.
Therefore, we employ explainable AI approaches to understand what tokens contribute to the generated code sequences.
We expect our exploratory experiments on explainable automated program generation can put forward practical insights for future research.
We employ a state-of-the-art model-agnostic explainable approach for interpreting language models.
We then utilize the explanation results to understand why program generation models output new code sequences effectively or ineffectively.



\section{Result}
\label{sec:result}
To answer the aforementioned RQs, we perform an extensive analysis of the generated code sequences and their explanation results, respectively.
Below, we present the results with respect to our three research questions.

\begin{table}[t]
  \centering
  \normalsize
  \caption{Performance of language models on the studied program generation datasets (accuracy)
  }
  \resizebox{1.01\linewidth}{!}{
    \begin{tabular}{lrrrrrrrrrrrr}
    \toprule
          & \textbf{Android\_S} & \textbf{Android\_M} & \textbf{Google\_S} & \textbf{Google\_M} & \textbf{Ovirt\_S} & \textbf{Ovirt\_M} & \textbf{CodeReview} & \textbf{B2F\_S} & \textbf{B2F\_M} & \textbf{Java2C\#} & \textbf{C\#2Java} & \textbf{CONCODE} \\
    \midrule
    \textbf{T5} & 6.42\% & 2.75\% & 4.93\% & 1.11\% & 16.85\% & 9.65\% & 18.00\% & 15.15\% & 5.58\% & 54.40\% & 62.40\% & 20.05\% \\
    \textbf{CoTexT} & 7.50\% & 3.98\% & 5.20\% & 1.80\% & 16.64\% & 10.99\% & 20.18\% & 16.61\% & 6.40\% & 57.00\% & 65.70\% & 20.70\% \\
    \textbf{CodeTrans} & 7.02\% & 3.30\% & 4.23\% & 0.97\% & 11.96\% & 7.12\% & 8.88\% & 8.14\% & 2.60\% & 49.10\% & 52.10\% & 21.65\% \\
    \textbf{CodeBERT} & 8.50\% & 5.01\% & 3.17\% & 0.83\% & 14.64\% & 10.13\% & 21.03\% & 12.10\% & 4.19\% & 50.20\% & 54.70\% & 18.60\% \\
    \textbf{CodeGPT} & 11.37\% & 8.14\% & 8.99\% & 4.50\% & 19.92\% & 12.05\% & 16.70\% & 13.61\% & 4.64\% & 59.90\% & 64.50\% & 17.65\% \\
    \textbf{CodeT5} & 13.80\% & 9.21\% & 10.40\% & 4.82\% & 22.06\% & 12.64\% & 28.31\% & 17.60\% & 8.34\% & \cellcolor[rgb]{ .996,  .878,  .824}64.10\% & 69.90\% & 22.35\% \\
    \textbf{CodeReviewer} & 14.68\% & 10.40\% & 11.81\% & 6.85\% & 25.49\% & 18.18\% & \cellcolor[rgb]{ .996,  .878,  .824}30.43\% & 17.94\% & \cellcolor[rgb]{ .996,  .878,  .824}8.77\% & 63.10\% & 70.40\% & \cellcolor[rgb]{ .996,  .878,  .824}22.65\% \\
    \textbf{CodeT5+} & \cellcolor[rgb]{ .996,  .878,  .824}15.20\% & \cellcolor[rgb]{ .996,  .878,  .824}11.36\% & \cellcolor[rgb]{ .996,  .878,  .824}14.27\% & \cellcolor[rgb]{ .996,  .878,  .824}7.27\% & \cellcolor[rgb]{ .996,  .878,  .824}26.06\% & \cellcolor[rgb]{ .996,  .878,  .824}20.03\% & 30.12\% & \cellcolor[rgb]{ .996,  .878,  .824}18.44\% & 7.84\% & 63.90\% & \cellcolor[rgb]{ .996,  .878,  .824}70.60\% & 21.85\% \\
    \bottomrule
    \end{tabular}%
    }
  \label{tab:tab1_performance_compare}%
\end{table}%

\subsection{(RQ1) How do language models perform on program generation tasks?}
\smallsection{Approach}
To better understand how well the pre-trained language models are in automated program generation, we extensively study and compare their performance on the studied datasets.
We follow the same training/validation/testing data splits and fine-tuning procedure for each dataset introduced in Section~\ref{sec:dataset}.
We employed a beam search setting of one for our evaluations. 
Then, we calculated the accuracy (i.e., the exact match rate) for each model. Table~\ref{tab:tab1_performance_compare} presents a detailed summary of the accuracy achieved by eight different language models across these datasets.

\smallsection{Result}
We observe from Table~\ref{tab:tab1_performance_compare} several key findings on language model performance on program generation tasks.
First, we find that language models exclusively pre-trained on natural language corpora, such as T5, demonstrate limited effectiveness on programming tasks compared to models pre-trained on code data.
For example, T5 achieves just 6.42\% and 2.75\% accuracy on \textit{Android\_S} and \textit{Android\_M} respectively, highlighting the importance of domain-specific pre-training in improving performance on automated program generation.
Additionally, we observe that model architecture plays a significant role in performance. 
Decoder-only (CodeGPT) and encoder-only (CodeBERT) models exhibit inferior results across multiple datasets compared to encoder-decoder architectures like CodeT5. 
Furthermore, we observe that models such as CodeT5, CodeReviewer, and CodeT5+ consistently outperform others across various benchmarks. 
For example, on the code+comment-to-code \textit{CodeReview} dataset, CodeReviewer obtains state-of-the-art performance with 30.43\% accuracy. Similarly, it leads text-to-code performance on the 
\textit{CONCODE} dataset with 22.65\% accuracy.
In other code-to-code scenarios, like the Tufano~\ea benchmarks, CodeT5+ performs better on multiple datasets, e.g., achieving 11.36\% accuracy on \textit{Android\_M}.
These findings are in alignment with previous studies, confirming the superiority of these models in program generation tasks~\cite{li2022codereviewer, hong2022commentfinder, wang2023codet5+}.

However, a critical observation is that the overall accuracy levels across all models are relatively low, especially in more complex tasks, which raises significant concerns about their reliability in real-world settings. 
When comparing dataset sizes, it is evident that all models perform better on small-sized datasets than on medium-sized ones. 
For example, CodeT5+'s accuracy fluctuates markedly from 18.44\% on the small \textit{B2F\_S} dataset down to just 7.84\% on the medium \textit{B2F\_M} dataset.
Additionally, the performance gap across different datasets is striking, as seen with CodeT5+, which achieves a high of 70.6\% on \textit{C\#2Java} but decreases to a mere 7.27\% on \textit{Google\_M}. 
Such fluctuations and inconsistency in results pose substantial challenges in understanding why models perform so well or poorly on specific datasets.
Understanding the reasons is crucial for future research, particularly for improving the reliability and practical application of automated program generation.

\find{\textbf{Finding 1:}
Pre-trained encoder-decoder-based models like CodeReviewer and CodeT5+ show superior performance compared to other models across diverse datasets. 
However, their inconsistent accuracy and significant variability across different tasks and datasets raise concerns about their reliability, emphasizing the need for a deeper understanding of factors affecting model behavior in downstream tasks.  
}

\subsection{(RQ2) How reliable are automated program generation approaches?}
Reliability and trustworthiness are critically important in automated program generation, especially within the evolving landscape of software engineering~\cite{lo2023trustworthy}.
Our preliminary investigation (RQ1) revealed a large disparity in the performance of language models: while some models demonstrated exceptionally high performance on certain datasets yet exhibited markedly lower effectiveness on others. 
This considerable fluctuation raises concerns about the potential overestimation or underestimation of their capabilities due to experimental biases.
Considering the potential for both overconfidence and undue skepticism resulting from these inaccuracies, it is important to measure these tools' effectiveness accurately, ensuring their proper application and deployment in software engineering practices.
In response to these issues, we undertake an in-depth analysis to uncover potential sources of unrealistic performance evaluation.
Our analysis is structured along three different aspects:
\begin{itemize}
    \item \textbf{Data Duplication between Training and Testing Sets.} 
    We investigate how similarities or duplications within training and testing datasets might inflate the perceived performance of language models. 
    Such overlaps may create an illusion of high accuracy, masking the true capabilities of these models in novel or diverse scenarios.
    \item \textbf{Data Duplication across Testing Sets.} 
    We investigate the presence of duplicate examples within testing datasets.
    Duplication within these sets can lead to a misleading evaluation.
    \item \textbf{Output-Input Similarity Analysis.} Finally, we examine the correlation between the outputs generated by the models and their inputs. In automated program generation, the expectation is for models to update or refine input code creatively and accurately. However, when outputs are the same as inputs, it raises questions about the true generative capacity of the models.
\end{itemize}

In our analysis, we utilize the BLEU score, specifically the \textit{BLEU-4} variant, to systematically evaluate the similarity within the studied datasets.
\textit{BLEU-4}, which assesses the co-occurrence of 4-gram sequences, is widely used in prior research~\cite {zeng2022extensive, Patanamon2022AutoTransform, hou2023large}.
A \textit{BLEU-4} score of 0 corresponds to no similarity, indicating completely unique content, while a score of 1 reflects total duplication or exact replication. 
By applying this metric, we can discern the extent to which our datasets contain unique or duplicative examples, thereby providing an empirical basis for evaluating the potential impact of data duplication on evaluation performance.

\subsubsection{Data Duplication between Training and Testing Sets}
In automated program generation, the robustness of language model evaluations is crucial. 
A key threat to this robustness is 'data snooping', a pitfall where models, due to improper data handling, gain inadvertent access to testing information during training~\cite{she2023pitfalls}. 
Such exposure can lead to exaggerated performance metrics as models may simply recall information rather than apply learned patterns to new data. 
To prevent this and ensure genuine model generalization, it is essential to assess the overlap between training and testing datasets.

\smallsection{Approach}
To determine the similarity score for a test instance $t$, we compare its source sequence against each instance in the training set $T = \{t_1, t_2, ..., t_n\}$. The similarity score, $S(t)$, is the maximum BLEU-4 score obtained from these comparisons:

\begin{equation}
S(t) = \max_{t_i \in T} \text{BLEU-4}(t, t_i)
\end{equation}

Additionally, we keep track of the index $i$ that yields this maximum score, which can be defined as:

\begin{equation}
i_{\text{max}}(t) = \argmax_{t_i \in T} \text{BLEU-4}(t, t_i)
\end{equation}

This iterative process is performed for each test instance $t$, allowing us to quantify the extent of data overlap and identify the potential duplication within the datasets.

\begin{figure}[t]
  \centering
  \includegraphics[width=1.02\textwidth]{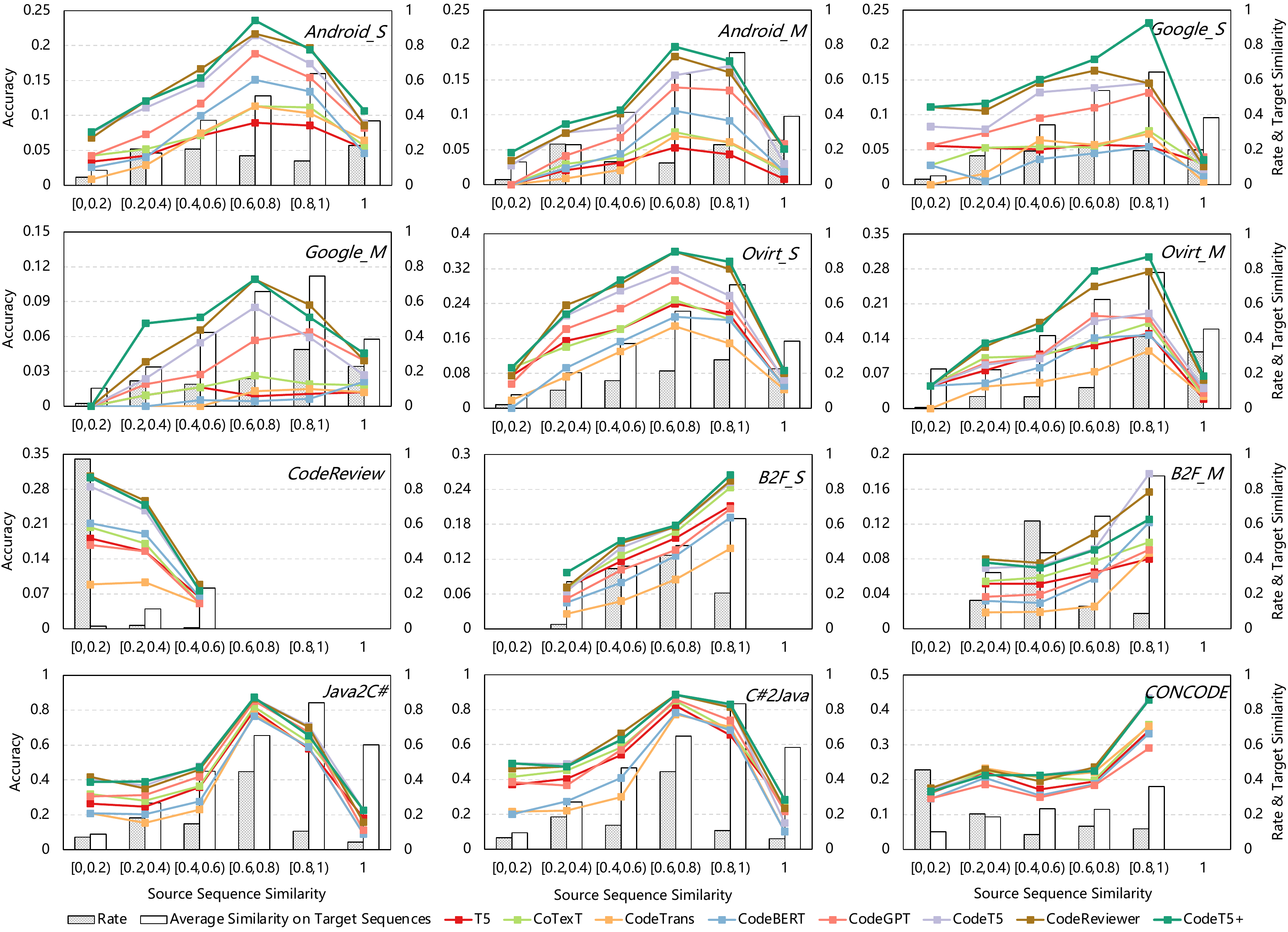}
  \caption{Distribution of Test-Training Data Duplication and Model Performance across Dataset
  }
  \label{fig:fig_train_test_duplicates}
\end{figure}

\smallsection{Result}
Figure~\ref{fig:fig_train_test_duplicates} and Table~\ref{tab:tab2_duplication_changes_codet5} present a detailed analysis of the data similarity between training and testing datasets and its influence on model performance.
Figure~\ref{fig:fig_train_test_duplicates} presents the distribution of test data similarity to training data across various datasets and corresponding recalculations of model performance for each similarity range.
Furthermore, we have recorded the average similarity score of the output sequences for each test instance with their most similar training instances, as shown in Figure~\ref{fig:fig_train_test_duplicates}.

Our analysis uncovers a wide variation in the similarity scores across an array of datasets, with the notable exception of the \textit{CodeReview} dataset, which stands out due to its lack of highly similar instances.
However, datasets such as \textit{Ovirt\_M} show that a majority, exceeding 60\%, of test data source sequences have a similarity score above 0.8, highlighting a considerable overlap with the training data. 
In particular, from \textit{Tufano et al.}, comprising \textit{Android\_S}, \textit{Android\_M}, \textit{Google\_S}, \textit{Google\_M}, \textit{Ovirt\_S}, and \textit{Ovirt\_M}, along with the \textit{CodeTrans-Dataset} datasets \textit{Java2C\#} and \textit{C\#2Java}, we find a significant presence of test samples that are duplicates of the training set (i.e, similarity score = 1).
For example, in all \textit{Tufano et al.} datasets, more than 20\% of the test samples are identical to those in the training set.
These findings suggest the presence of data duplication within the datasets, raising important questions about the possibility of data snooping that could distort the evaluation of model performance.

In Figure~\ref{fig:fig_train_test_duplicates}, we observe an increase in the similarity of target sequences corresponding to an increase in source sequence similarity, up until the source are completely identical ($S(t)=1$). 
Taking the \textit{Android\_S} dataset as an example, the average similarity for target sequences climbs from 0.1 to approximately 0.6 as the similarity of the source sequences increases. 
This indicates that target sequences are generally more aligned when the sources are more similar.
However, when the source sequences are identical ($S(t)=1$), the similarity between target sequences notably drops.
For example, in \textit{Android\_S}, the average similarity on target sequences drops from 0.6 to 0.4.
This observation may indicate a research oversight during dataset preparation, where instances with identical "source + target" pairs are usually removed, but those with identical sources or targets, when considered separately, remain unfiltered.

Figure~\ref{fig:fig_train_test_duplicates} illustrates that as the similarity between test and training instances increases, so does the performance of language models. 
For instance, CodeT5+ show an accuracy of around 10\% for test instances with a similarity score below 0.2, which increases to nearly 25\% for instances with a similarity score above 0.8 in the \textit{B2F\_S} dataset. 
As shown in Table~\ref{tab:tab1_performance_compare}, the average performance of CodeT5+ on the \textit{B2F\_S} dataset is 18.44\%.
It suggests that models may be leveraging memorized patterns from the training data rather than demonstrating true generalization capabilities.
Consequently, the ability of language models to handle unseen or novel test instances remains a considerable challenge for future research.
A notable drop in model performance is also observed when the source sequences are exactly the same, a decrease that corresponds with the fall in similarity among output sequences. 
This could be due to the common practice during dataset preparation where instances of identical "source + target" pairs are removed, possibly leading to an oversight of singular duplicates in sources or targets.

Table~\ref{tab:tab2_duplication_changes_codet5} further describes the impact on the performance of models such as CodeReviewer and CodeT5+ when instances with high similarity scores are excluded from the test sets. 
While the \textit{CodeReview} dataset shows no change due to a lack of closely matched samples, other datasets usually display a marked decrease in accuracy once we remove test cases with high similarity scores.
In the \textit{Android\_S} dataset, for example, the accuracy of \textit{CodeT5+} decreases from 15.2\% to 13.09\%. In some cases, the drop is even more apparent; \textit{CodeT5+} falls from 70.6\% to 53.09\% on the \textit{C\#2Java} dataset.
This highlights the critical role that a varied and comprehensive test set plays in ensuring the accuracy of a model's performance evaluation.

\find{\textbf{Finding 2:}
Our results reveal that multiple program generation datasets contain substantial duplications between their training and testing sets. 
The increased similarity between these sets typically leads to exaggerated performance metrics, raising concerns about the models' generalization capabilities and suggesting potential flaws in data handling.}

\begin{table}[t]
  \centering
  \large
  \caption{Model Performance Before and After Removing High-Similarity Test Instances.}
  \resizebox{\linewidth}{!}{  
    \begin{tabular}{clrrrrrrrrrrrr}
    \toprule
          &       & \textbf{Android\_S} & \textbf{Android\_M} & \textbf{Google\_S} & \textbf{Google\_M} & \textbf{Ovirt\_S} & \textbf{Ovirt\_M} & \textbf{CodeReview} & \textbf{B2F\_S} & \textbf{B2F\_M} & \textbf{Java2C\#} & \textbf{C\#2Java} & \textbf{CONCODE} \\
    \midrule
    \multicolumn{2}{r}{\textbf{Test Samples Percentage (>0.6)}} & 53.69\% & 60.62\% & 60.88\% & 71.21\% & 71.72\% & 85.74\% & 0.05\% & 62.81\% & 21.82\% & 59.80\% & 61.20\% & 25.25\% \\
    \midrule
    \multirow{4}[2]{*}{\textbf{CodeReviewer}} & \textbf{Original Accuracy} & 14.68\% & 10.40\% & 11.81\% & 6.85\% & 25.49\% & 18.18\% & 30.43\% & 17.94\% & 8.77\% & 63.10\% & 70.40\% & 22.65\% \\
          & \textbf{New Accuracy} & 13.61\% & 8.02\% & 12.61\% & 4.81\% & 25.25\% & 14.39\% & 30.44\% & 14.24\% & 7.64\% & 40.30\% & 53.87\% & 19.26\% \\
          & \textbf{Original BLEU} & 0.70  & 0.72  & 0.71  & 0.73  & 0.75  & 0.77  & 0.86  & 0.75  & 0.85  & 0.92  & 0.93  & 0.59 \\
          & \textbf{New BLEU} & 0.70  & 0.72  & 0.70  & 0.67  & 0.73  & 0.72  & 0.86  & 0.76  & 0.85  & 0.89  & 0.91  & 0.56 \\
    \midrule
    \multirow{4}[1]{*}{\textbf{CodeT5+}} & \textbf{Original Accuracy} & 15.20\% & 11.36\% & 14.27\% & 7.27\% & 26.06\% & 20.03\% & 30.12\% & 18.44\% & 7.84\% & 63.90\% & 70.60\% & 21.85\% \\
          & \textbf{New Accuracy} & 13.09\% & 9.07\% & 13.29\% & 6.97\% & 25.13\% & 14.21\% & 30.14\% & 14.75\% & 7.11\% & 42.04\% & 53.09\% & 18.39\% \\
          & \textbf{Original BLEU} & 0.70  & 0.72  & 0.72  & 0.72  & 0.75  & 0.77  & 0.85  & 0.75  & 0.85  & 0.93  & 0.93  & 0.59 \\
          & \textbf{New BLEU} & 0.70  & 0.72  & 0.71  & 0.66  & 0.74  & 0.71  & 0.85  & 0.76  & 0.85  & 0.89  & 0.91  & 0.56 \\
    \bottomrule
    \end{tabular}%
   }
  \label{tab:addlabel}%
  \label{tab:tab2_duplication_changes_codet5}
\end{table}%

\subsubsection{Data Duplication across Testing Sets}
Building upon the test-training analysis, this section examines intra-set duplications within the test datasets.
The presence of duplicated instances in test sets can lead to unreliable evaluation results.
If such duplications go unaddressed, the model's efficacy may end up being evaluated on a narrow subset of repeated test instances rather than a diverse range of samples.
To Address this concern, we investigate the prevalence of data duplication within the testing sets based on our studied datasets.

\smallsection{Approach}
For each test instance $t$, we investigate potential duplication within the test set by comparing its source sequence with that of every other test instance. The maximum BLEU-4 score from these comparisons is recorded, along with the index of the test instance that yields this maximum score:

\begin{equation}
D(t) = \max_{\substack{t' \in \mathbf{T}_{\text{test}} \\ t' \neq t}} \text{BLEU-4}(t, t')
\end{equation}

\begin{equation}
j_{\text{max}}(t) = \argmax_{\substack{t' \in \mathbf{T}_{\text{test}} \\ t' \neq t}} \text{BLEU-4}(t, t')
\end{equation}

Here, $D(t)$ is the similarity score for the test instance $t$ within testing data, and $j_{\text{max}}(t)$ is the index of another test instance $t'$ that is most similar to $t$. This process is iteratively performed for each test instance in the dataset, allowing us to identify exact and near-duplicates within the test set.

\begin{figure}[t]
  \centering
  \includegraphics[width=1.02\textwidth]{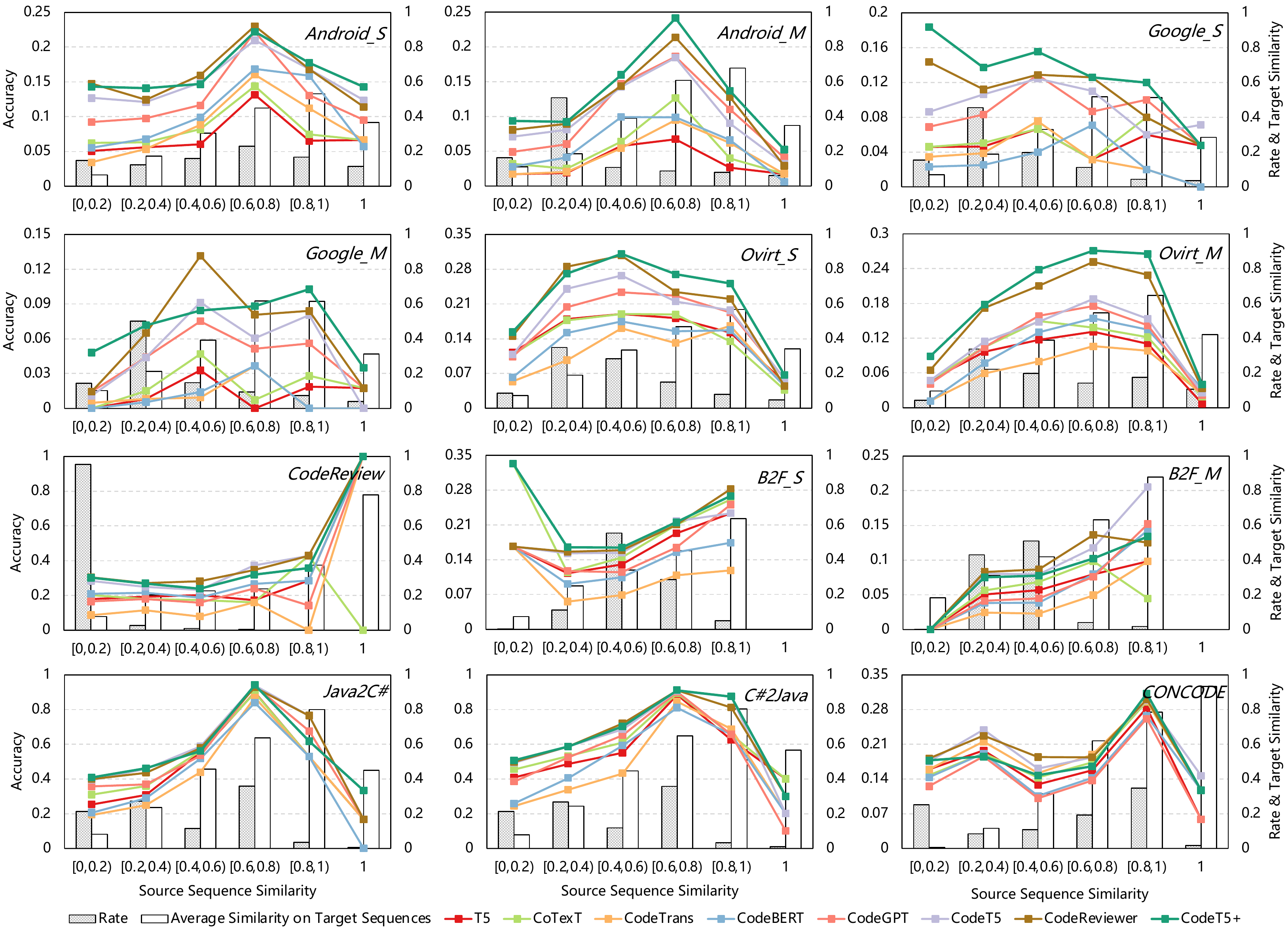}
  \caption{Distribution of Data Duplication within Testing Sets and Corresponding Model Performance Metrics
  }
  \label{fig:fig_within_test_duplicates}
\end{figure}

\begin{figure}[t]
  \centering
  \includegraphics[width=0.9\textwidth]{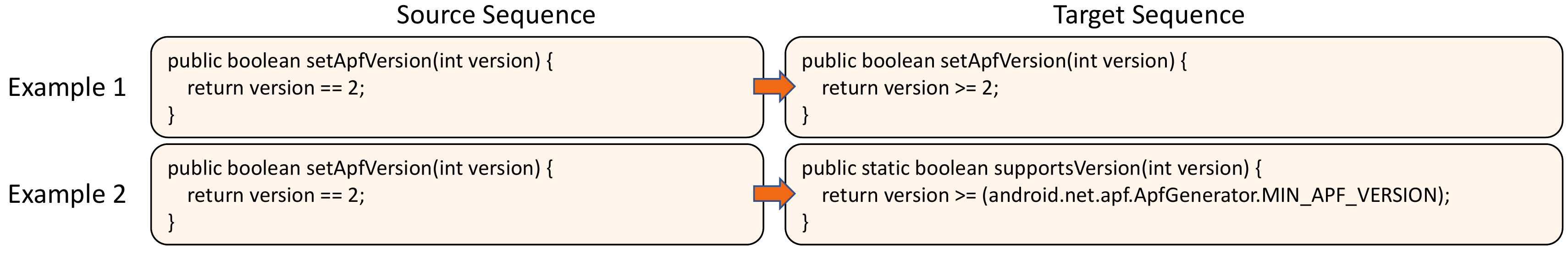}
  \caption{Examples of test instances with duplicated sources and different targets from \textit{Android\_S}.
  }
  \label{fig:fig_example_test_duplicates}
\end{figure}

\smallsection{Results}
Figure~\ref{fig:fig_within_test_duplicates} shows the distribution of similarity within the test data of our studied datasets. 
We observe that except for the \textit{B2F\_S} and \textit{B2F\_M} datasets, which exhibit no duplicate instances, a considerable number of datasets include duplicates. 
For example, \textit{Android\_S} records 11\% of its test instances as duplicates based on source sequences, and similarly, \textit{Ovirt\_M} possesses more than 10\% duplication.
In line with earlier observations from Figure~\ref{fig:fig_train_test_duplicates}, the average similarity scores of the target sequences for these duplicate instances are substantially lower compared to the dataset at large.
This indicates that while testing language model performance on program generation, identical source sequences are often provided, but the models are expected to provide different outputs.
Figure~\ref{fig:fig_example_test_duplicates} provides an example from the \textit{Android\_S} dataset where two test instances with the same source code require different correct outputs from the model. 
This practice leads to an inherently unfair evaluation scenario, where the same test instance is associated with different performance expectations. 
Figure~\ref{fig:fig_within_test_duplicates} indicates that performance on these duplicated test instances can vary greatly from the average, potentially giving an inaccurate representation of model performance.
Our findings emphasize the need to carefully construct test sets, avoiding the pitfalls of duplications that can compromise reliable model performance evaluation.
It's crucial to construct detailed and impartial evaluation approaches that truly measure the language models' ability to generate code across a wide array of test scenarios.

\find{\textbf{Finding 3:}
In our examination of 12 datasets, we find that 10 contain duplicated source sequences within their test instances, despite requiring models to generate different targets. Such inconsistencies in test design lead to evaluations that may not accurately reflect the true capabilities of program generation approaches, thus compromising the reliability of performance evaluation.
}

\subsubsection{Output-Input Similarity Analysis.}
The goal of automated program generation is to refine or create new code sequences that accurately address specified functionality changes or requirements.
In most prior works, model predictions are compared against target outputs to calculate accuracy and other performance metrics.
We extend this evaluation to compare the generated outputs with the original inputs, assessing whether the models are merely mirroring the inputs or actually generating updated code sequences. 

\smallsection{Approach}
Different from our earlier focus on similarity scores within and across test sets, here we focus specifically on the percentage of identical sequences in the model outputs as compared to their source code sequences. 
For this comparison, we utilize a direct string comparison method, discounting special tokens such as "<pad>", "<s>", and "</s>", which are typically used for formatting and do not contribute to the functional content of the code. 
Through this process, for each model's predictions, we calculate the proportion of outputs that are exactly the same as the source sequences.

\smallsection{Result}
Table~\ref{tab:output_input_duplicates} presents the rates of duplication between model outputs and inputs across several language models on the studied datasets. 
The table reveals a considerable variance in the rates across different datasets and models.
Notably, only the \textit{CodeReview}, \textit{Java2C\#}, and \textit{C\#2Java} datasets display minimal duplication between source and target sequences, with rates as low as 1-2\%, suggesting these datasets effectively evaluate the models' capacity for generating new code.
When comparing model outputs with source code sequences, language models present high rates of output-input duplication.
For example, the rate of T5 on the \textit{Android\_S} and \textit{Android\_M} reaches 78\% and 80\%, respectively.
Such a high duplication rate suggests that a substantial portion of the output code is replicated from the inputs, calling into question the models' generative capabilities.
Although models like CodeReviewer and CodeT5+ show superior performance, as indicated in RQ1, they still exhibit a significant degree of output duplication with their inputs (e.g., 35\% for CodeT5+ on the \textit{B2F\_S} dataset and 32\% on \textit{Google\_M}).
The analysis of the \textit{CodeReview} dataset, even within the context of code+comment-to-code tasks, reveals a significant level of duplication between the source code sequences and the models' outputs.
Note the \textit{CONCODE} dataset, which focuses on text-to-code tasks, demonstrates zero duplication, indicating the models' potential to generate entirely new code from textual prompts.
In the context of language models applied to code translation tasks within the \textit{Java2C\#} and \textit{C\#2Java} datasets, we observe a commendably low duplication rate. 
The observed variability in duplication rates across datasets and tasks underscores the imperative for nuanced and robust evaluation metrics that can accurately reflect the true generative capabilities of language models. 

\begin{table}[t]
  \centering
  \caption{Comparison of Output-Input Duplication Rates Across LMs for Program Generation Tasks}

  \resizebox{1\linewidth}{!}{
    \begin{tabular}{lrrrrrrrrrrrr}
    \toprule
          & \textbf{Android\_S} & \textbf{Android\_M} & \textbf{Google\_S} & \textbf{Google\_M} & \textbf{Ovirt\_S} & \textbf{Ovirt\_M} & \textbf{CodeReview} & \textbf{B2F\_S} & \textbf{B2F\_M} & \textbf{Java2C\#} & \textbf{C\#2Java} & \textbf{CONCODE} \\
    \midrule
    \textbf{Source vs Target} & 0\%   & 0\%   & 0\%   & 0\%   & 0\%   & 0\%   & 1\%   & 0\%   & 0\%   & 2\%   & 2\%   & 0\% \\
    \midrule
    \textbf{T5} & 78\%  & 80\%  & 83\%  & 81\%  & 53\%  & 61\%  & 37\%  & 21\%  & 71\%  & 2\%   & 2\%   & 0\% \\
    \textbf{CoTexT} & 74\%  & 73\%  & 80\%  & 77\%  & 54\%  & 58\%  & 33\%  & 22\%  & 73\%  & 2\%   & 2\%   & 0\% \\
    \textbf{CodeTrans} & 36\%  & 36\%  & 30\%  & 27\%  & 36\%  & 41\%  & 50\%  & 71\%  & 84\%  & 2\%   & 2\%   & 0\% \\
    \textbf{CodeBERT} & 21\%  & 27\%  & 10\%  & 5\%   & 18\%  & 29\%  & 36\%  & 51\%  & 78\%  & 1\%   & 1\%   & 0\% \\
    \textbf{CodeGPT} & 36\%  & 47\%  & 56\%  & 57\%  & 33\%  & 31\%  & 41\%  & 31\%  & 77\%  & 2\%   & 2\%   & 0\% \\
    \textbf{CodeT5} & 40\%  & 44\%  & 60\%  & 51\%  & 27\%  & 46\%  & 20\%  & 24\%  & 64\%  & 2\%   & 2\%   & 0\% \\
    \textbf{CodeReviewer} & 15\%  & 31\%  & 26\%  & 37\%  & 10\%  & 20\%  & 14\%  & 11\%  & 44\%  & 2\%   & 2\%   & 0\% \\
    \textbf{CodeT5+} & 21\%  & 29\%  & 31\%  & 32\%  & 16\%  & 18\%  & 17\%  & 10\%  & 35\%  & 2\%   & 2\%   & 0\% \\
    \bottomrule
    \end{tabular}%
    }
  \label{tab:output_input_duplicates}%
\end{table}%

\find{\textbf{Finding 4:}
In code review and code repair tasks, it is observed that language models frequently generate outputs that are identical to the input sequences, presenting a potential limitation in their ability to generate novel code solutions.
}

\subsection{(RQ3) Can we explain why automated program generation approaches can (or fail to) generate code sequences reliably?}
Most pre-trained language models operate as black-box systems, obscuring their internal decision-making processes.
As highlighted in RQ1, the accuracy of these models with beam search set at 1, is not particularly high, underscoring the uncertainty in their output reliability. 
Furthermore, the results from RQ2 suggest that the evaluation of these models may be compromised by certain impractical experimental settings. 
Consequently, it is necessary to employ explainable AI approaches to demystify the internal workings of these models.
In this section, we employ gradient-based SHAP to examine and understand the decision-making processes of automated program generation models. 
This approach aims to uncover the factors behind a model's ability or failure to generate accurate and reliable code sequences.
To ensure a focused and relevant analysis, we limit our examination to the models demonstrating superior performance in previous sections, i.e., CodeT5, CodeReviewer, and CodeT5+. 
This focused examination is designed to gain insights into what contributes to their effective and reliable performance in program generation, with the aim of guiding future developments in this field.

\subsubsection{Understanding Token Importance}
In this section, we delve into the learning patterns of language models concerning program generation by examining the average feature importance of different types of tokens.
These tokens represent the fundamental building blocks of programming languages and are crucial for understanding the model's focus during the program generation process. 
We analyze five types of tokens, including: 

\begin{itemize}
    \item \textbf{Identifiers} are unique names for variables, classes, methods, etc.
    \item \textbf{Keywords} refer to programming language-specific words with predetermined standard meanings.
    \item \textbf{Operators} are special symbols that represent some operation on data.
    \item \textbf{Separators} are special symbols used to indicate the group of code.
    \item \textbf{Literals} refer to constant values.
\end{itemize}

\begin{figure}[t]
  \centering
  \includegraphics[width=0.95\textwidth]{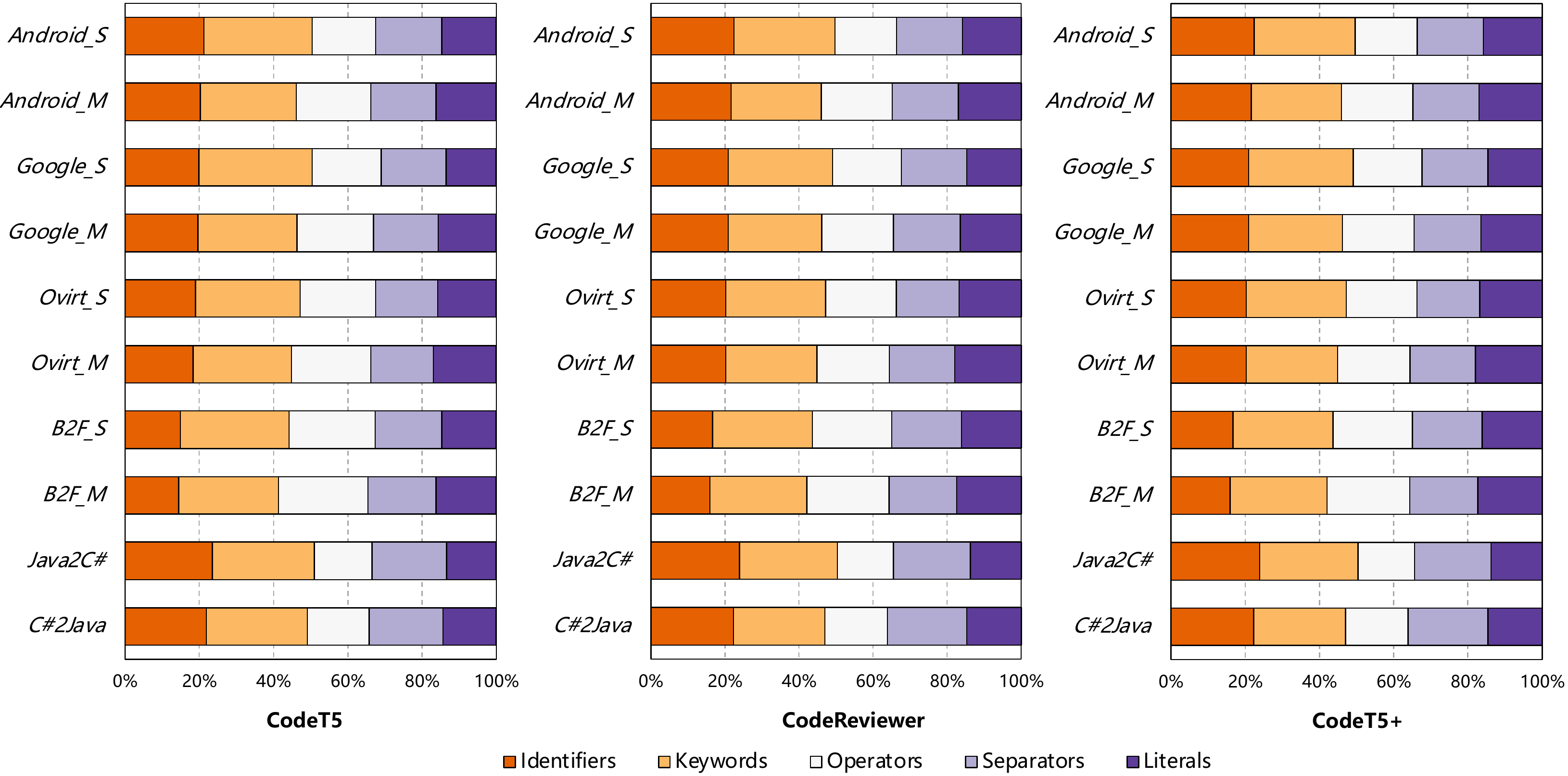}
  \caption{ Average feature importance of token types highlighted by language models on the studied datasets\protect\footnotemark. 
  }
  \label{fig:fig3_exp_types_different_models}
\end{figure}

\footnotetext{Note that the dataset \textit{CodeReview} and \textit{CONCODE} is excluded since it is challenging to distinguish the token types from the source sequences due to the abundance of comments and natural language descriptions.}

\smallsection{Results}
Figure~\ref{fig:fig3_exp_types_different_models} presents the average feature importance of token types across various datasets when analyzed through the CodeT5, CodeReviewer, and CodeT5+ models. 
The bar chart provides a comparative visualization of how each model weighs the significance of identifiers, keywords, operators, separators, and literals during the program generation process.
It is apparent across the models: Identifiers and keywords tend to be assigned higher importance scores, showing their critical role in understanding the syntactic and semantic structure of the code. 
This suggests that models are possibly prioritizing the recognition of variable names, function calls, and control structures, which are key for the functionality of the code.
Operators and separators, while varied across models and datasets, generally exhibit moderate importance. 
This reflects a slight comprehension by the models of the operational logic and structural delineation within the code, which, although less emphasized than identifiers and keywords, are still recognized as essential components of program logic.

\find{\textbf{Finding 5:}
The explanation results reveal that identifiers and keywords are consistently assigned higher importance scores compared to operators and separators, indicating that language models prioritize syntactic and semantic understanding in program generation.
}

\begin{figure}[t]
  \centering
  \includegraphics[width=0.7\textwidth]{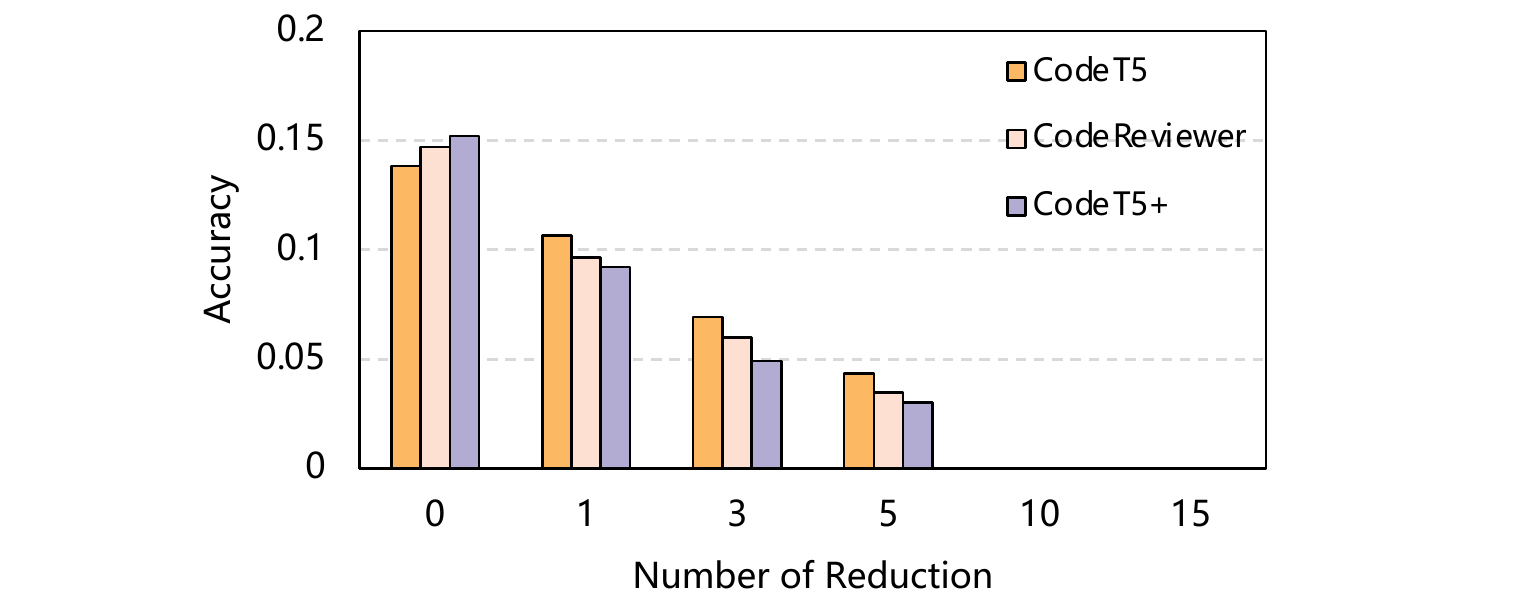}
  \caption{Impacts of token reduction size on \textit{Android\_S}
  }
  \label{fig:fig_exp_removal_numbers}
\end{figure}

\subsubsection{Model Performance Under Input Token Reduction}
In this section, we investigate the robustness of program generation models when faced with reduced input tokens.
Drawing from the explanation results, we investigate how the selective deletion of tokens classified as less important impacts the models' ability (i.e., lowest feature importance) to generate accurate code sequences.

\smallsection{Impacts of Token Reduction Size}
In this study, we examine the impact of selectively removing tokens from the input sequences based on their feature importance scores. For each test case, tokens are ranked by their importance, and we methodically remove a specified number of the least important tokens, with placeholders inserted to preserve the format of the code.
This process generates new "source-target" pairs, which are then fed back into the models. The experiment is conducted with incremental token removals, specifically at counts of 1, 3, 5, 10 and 15, to understand how the absence of certain tokens affects the models' code generation capabilities. 
In our analysis, we denote the original "source-target" testing instances with a 0 to differentiate them from the modified instances.

\smallsection{Results}
The results presented in Figure~\ref{fig:fig_exp_removal_numbers}, which focus on the \textit{Android\_S} dataset, provide a clear illustration of the performance degradation associated with the incremental removal of the least important tokens. 
For example, in the CodeT5+ model, eliminating just one token leads to a significant drop in performance, from 15.2\% to 9.19\%.
As the number of tokens removed increases to five, the performance further declines to a mere 3\%. 
Beyond the removal of ten tokens, the model's performance drops to zero, indicating a complete inability to generate the correct code sequence.
This pattern is not unique to CodeT5+; similar trends are observed across all three studied models.
The significant decrease in performance, even with the removal of tokens previously considered less important, suggests a complex interdependence among the various input features. 
These findings raise important questions about the robustness of program generation models, particularly their sensitivity to changes in input and their ability to adapt and maintain accuracy under modified conditions.

\begin{figure}[t]
  \centering
  \includegraphics[width=\textwidth]{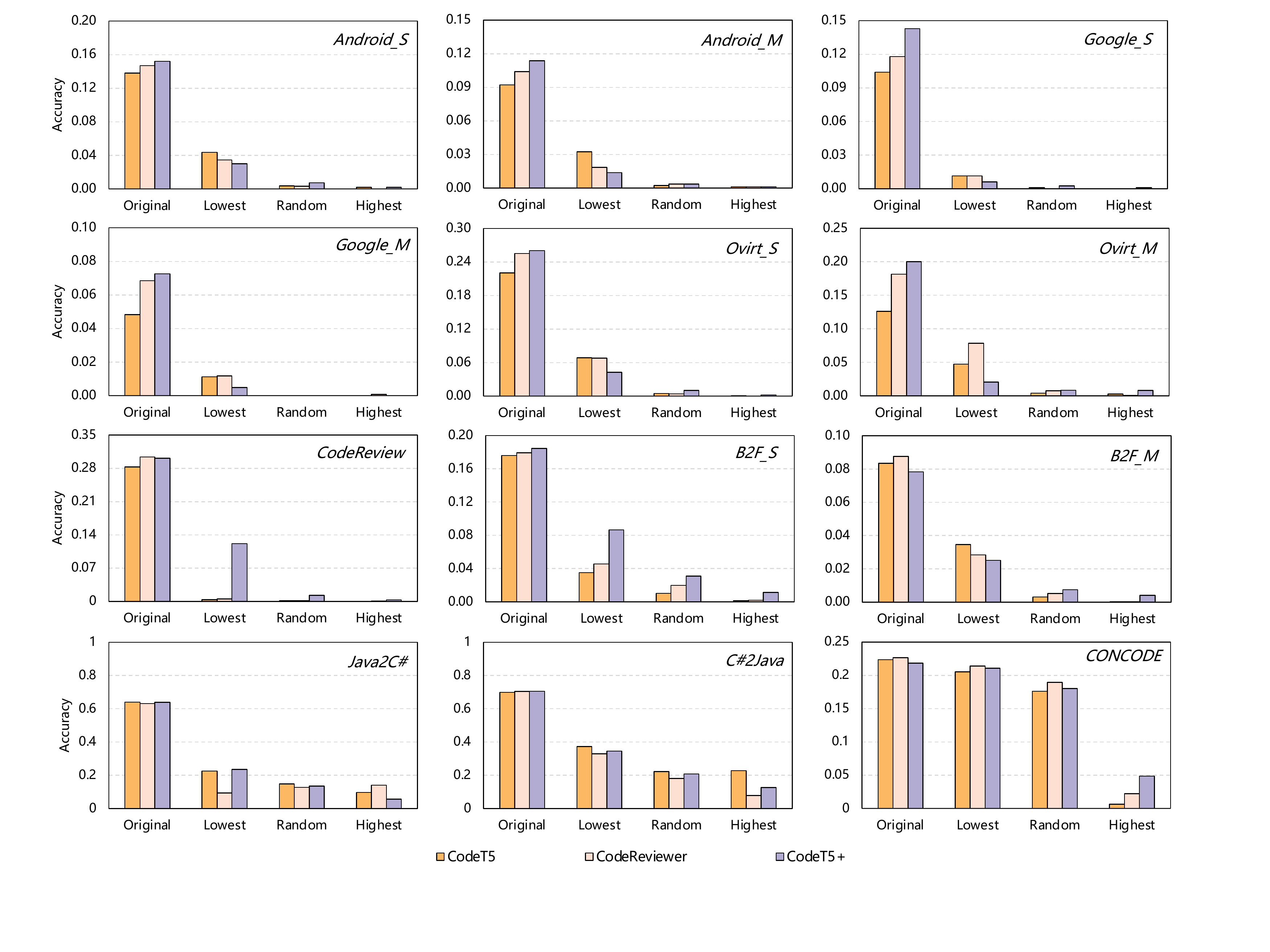}
  \caption{Performance upon input token reduction using different strategies
  }
  \label{fig:fig_exp_removal_different_str}
\end{figure}

\smallsection{Comparative Analysis of Token Removal Strategies}
To evaluate the reliability of the explainable AI approach, we compare strategies to remove tokens from the input sequences, specifically targeting the lowest and highest importance tokens identified by our model, and also using a random token removal for comparison.
This comparison is designed to investigate how each removal strategy affects the model's performance. By comparing the results of targeted versus random token removal, we aim to examine the accuracy of the explainable AI approach's ability to identify important features for program generation. 
In each of these three strategies, we consistently remove a set number of five tokens for a standardized comparison.

\smallsection{Results}
As shown in Figure~\ref{fig:fig_exp_removal_different_str}, the removal of tokens demonstrates a clear impact on performance across different datasets. 
Specifically, removing tokens identified with the lowest importance typically leads to a smaller decline in performance compared to either random removal or the removal of the most important tokens. 
For example, within the \textit{Android\_S} dataset, the accuracy of the CodeT5+ model falls to 3\% after removing the least important tokens, but drops almost to zero with random or most important token removal. 
Similarly, within the \textit{B2F\_S} and \textit{CONCODE} dataset, the accuracy of the CodeT5+ model on lowest and random removal is much higher compared to removing the most important tokens. 
Thus, these findings suggest that explainable AI approaches can help us to determine the significance of different input tokens for program generation. 
Moreover, the consistent decrease in performance across various datasets, following the removal of important tokens, presents a lack of robustness in these language models. 

\find{\textbf{Finding 6:}
Our results show that explainable AI methods can effectively identify feature importance when generating code sequences.
The substantial performance decrease observed when crucial tokens are removed underlines a vulnerability in language models, showing a need for enhanced model robustness.
}

\section{Discussion}
\label{sec:discussion}
In this study, we have identified several interesting findings of the reliability and explainability of automated program generation approaches. 
We now discuss the main implications and limitations of our study.

\subsection{Implications}

\noindent\textbf{\underline{Reliability.}}
High reliability is essential for the real-world usage of language model-based automated program generation systems. 
As highlighted by prior works~\cite{yang2022natural, schuster2021you}, state-of-the-art language models are vulnerable to adversarial attacks and can be fooled into recommending wrong code.
This finding indicates that the existing language models still suffer from reliability issues.
However, most prior works pay more attention to improving the accuracy of automated program generation systems, but neglect to evaluate whether the proposed methodologies are sufficiently reliable.

Our study first replicates the good performance presented in previous research.
Our results further indicate that data duplication commonly exists in existing state-of-the-art program generation datasets (i.e., duplication between training and testing sets, as well as duplication within testing sets), and the issues provide unreliable or unrealistic performance evaluation in prior research. 
Further, our results show that pre-trained language models frequently generate outputs replicated from the inputs, outputting numerous unchanged code sequences.
These findings provide evidence that LM-based program generation approaches suffer from serious reliability threats since the performance is overestimated or underestimated by unreliable experimental analysis.
This not only provides unreliable evaluation performance of DL systems, but also raises concerns regarding their deployment in real-world applications.
Consequently, there is a need for research on enhancing the quality and reliability of automated program generation research, which can further benefit the deployment of DL systems in the real world. 
First, more research efforts should focus on the dataset quality. 
Except for the data duplication issues~\cite{allamanis2019adverse}, Shi~\ea~\cite{shi2022we} and Sun~\ea~\cite{sun2022importance} have demonstrated that data noise was prevalent in widely-used benchmark datasets in code summarization and code search. 
Therefore, for future research, it is important to construct a rigorous evaluation methodology supported by reliable and standardized benchmarks.
Also, instead of only relying on accuracy, more comprehensive evaluation metrics should be employed to provide a reliable evaluation for language models.

\noindent\textbf{\underline{Explainability.}}
Pre-trained models are black-box models.
Thus, prior research usually disregards understanding why the LM-based program generation approaches make a specific prediction.
Our study employs a model-agnostic explainable AI approach to explain language models.
Furthermore, our results reveal several insightful findings that can inspire future research.
First, we observe that LM-based program generation models pay much more attention to keywords and identifiers of programming language compared with operators and separators.
This indicates that language models are capable of recognizing code grammar and structural information. 
However, our findings reveal a significant decline in the performance of language models with the removal of even a few tokens, including those considered least important, highlighting a lack of robustness in these models.
These observations from explanation results help me better understand inference behaviors and learning abilities of program generation behaviors. 
Consequently, our study also proves that explainable AI approaches are an effective and promising approach to analyzing or improving the reliability of LM-based automated program generation systems.
More research can be initiated on this aspect.

\subsection{Threats to Validity}
The primary threat to internal validity mainly lies in the model architecture and hyper-parameter setting.
We use eight program generation models, which are based on the same model settings in the original papers.
It is expected that hyper-parameter tuning would bring performance improvements.
Instead, the goal of our work is not to find the best setting, but to fairly investigate the reliability and explainability of program generation models.
The external threats to validity mainly lie in the studied datasets and the generalizability of the results.
In this study, we used five different types of program generation datasets.
We considered four types of code generation tasks (i.e., code review, code repair, code translation, and code generation), different token sizes (i.e., small and medium), different programming languages (e.g., java, python, C\#), and three types of inputs (i.g., only code, code+comments, only text). 
For reproducibility purposes, we provided a replication package to facilitate future work to replicate our approach on more repositories and tasks.
\section{Related work}
\label{sec:related_work}
\subsection{Language Models for Program Generation}
In recent years, researchers have increasingly applied language models, including pre-trained models, to program generation tasks.
Tofuno~\ea~\cite{tufano2019learning} develops a deep learning approach using RNNs to automatically transform source code in the code review context.
Thongtanunam~\ea~\cite{Patanamon2022AutoTransform} further introduces advanced Transformer architecture and a Byte-Pair Encoding (BPE) approach to handle the Out-Of-Vocabulary and long sequence problems. 
To better learn code properties, pre-training techniques are increasingly adopted in the code review scenario~\cite{li2022codereviewer,zhang2022coditt5,chirkova2021empirical, hong2022commentfinder, Tufano2022Using}. 
Hong~\ea~\cite{hong2022commentfinder} proposes a CodeT5-based approach to recommend code review comments automatically.
Li~\ea~\cite{li2022codereviewer} develops a transformer-based encoder-decoder model that is pre-trained on large code reviewer-specific data for code refinement tasks.
Except that, many studies~\cite{chen2018tree, lachaux2020unsupervised} involve translating code from one programming language to another. 

In contrast to previous research, this paper undertakes an empirical study to evaluate how reliable language models are in program generation. 
Most prior studies demonstrate that the proposed approach is more accurate,
However, these studies have not been able to show why it is more accurate convincingly.
Our study focuses on understanding the mechanisms behind language models in program generation scenarios and attempting to analyze potential issues concealed by decent performance. 

\subsection{Explainable LM-based Program Generation}
Previous research has already investigated applying explainable AI approaches, despite not being well-suited, to automated program generation~\cite{ tantithamthavorn21xai, fu2022LineVul, Liu2022XAI, Jiarpakdee22XAI, tantithamthavorn21actionable}.
There have been many works that explore the attention mechanism to explain language models of code~\cite{zhang2022diet, ahmad2021unified, Mohammadkhani2023, paltenghi2021thinking, wan2022they}.
Zhang~\ea~\cite{zhang2022diet} employed attention mechanisms to dig into critical statements and tokens learned by pre-trained language models in code search and code summarization. 
Paltenghi~\ea~\cite{paltenghi2021thinking} explored to what extent the attention weights of language models match the reasoning of skilled humans in code summarization.
Wan~\ea~\cite{wan2022they} analyzed the self-attention weights of pre-trained models of code and discovered that attention could capture high-level source code structural information.
Besides the attention mechanism, various explainable AI approaches have also been employed to explain NMT models of source code.
For example, Cito~\ea~\cite{cito2022counterfactual} integrated counterfactual explanation techniques for language models that predict certain properties of code or code changes. 
Rabin~\ea~\cite{rabin2021understanding} provided a model-agnostic approach to identify critical input features for models of code and demonstrated that the approach enables code simplification in code search and variable misuse debugging. 

Prior work has mainly focused on explaining code-to-text or text-to-code tasks (e.g., code search and code summarization).
Our study aims to understand the programming language models for program generation (i.e., code-to-code). 
In addition, previous works capture structural knowledge learned by language models, while we dig more into model behavior understanding in various experimental scenarios.

\subsection{Robustness of Language Models for Source Code}
Owing to the increasing research interest in this pattern, there are many recent studies proposed to explore the robustness and effectiveness of source code models. 
A plethora of studies have demonstrated that source code models are vulnerable to adversarial attacks~\cite{schuster2021you, yang2022natural, nguyen2021adversarial, zhou2022adversarial}. 
For example, Yang~\ea~\cite{yang2022natural} developed ALERT, a black-box attack approach that adversarially modifies code snippets to force pre-trained code models to produce incorrect outputs.
Schuster~\ea~\cite{schuster2021you} showed that code completion models are susceptible to poisoning attacks by adding some carefully crafted files to the training data of a model. 
Some empirical studies have been conducted to empirically investigate the performance of programming language models~\cite{zeng2022extensive, mastropaolo2021studying, Ciniselli2021Empirical, wang2022no}.
For example, Zeng~\ea~\cite{zeng2022extensive} suggests that developing an almighty pre-trained code model across task types is challenging, and more rigorous evaluations are required.

Distinct from previous work, we empirically examine the robustness and limitations of pre-trained code models on automated program generation.
Furthermore, we apply explainable AI approaches to assist us in understanding why code models are not robust enough for automated language generation, which is currently rarely explored in software engineering.

\subsection{Reliability in Language Models for Source Code}
With the increasing application of advanced large language models in software engineering, ensuring their trustworthiness becomes important~\cite{lo2023trustworthy, liu2022autoupdate}. 
She~\ea~\cite{she2023pitfalls} have reviewed common experimental biases in language models for code research, such as data noise, labeling errors, and inappropriate evaluation approaches. 
Allamanis~\ea~\cite{allamanis2019adverse} have investigated the effects of code duplication and they found that performance metrics could be inflated by up to 100\% when testing on duplicated code corpora, as opposed to de-duplicated corpora which more accurately reflect real-world usage by software engineers. 
Nie~\ea~\cite{nie2023understanding} explored labeling errors in vulnerability detection datasets, noting that incorrectly labeling a non-vulnerable sample as vulnerable was a more common issue.   
Additionally, Nong~\ea~\cite{nong_generating_2022} highlighted that datasets like SARD~\cite{black2017sard} often contain synthetic examples that are not representative of real-world code, characterized by smaller vocabulary, shorter program lengths, and higher pattern frequency.
As language models for code begin to be deployed in real-world applications (e.g., GitHub Copilot~\cite{github2021copilot} and ChatGPT~\cite{openai2023gpt4}), new challenges emerge, such as real-world constraints, security concerns, and vulnerability to attacks~\cite{zhang2023practices, mastropaolo2023robustness, dakhel2023github, liu2023refining}.

Different from prior studies, our research specifically focuses on reliability issues in program generation. We comprehensively examine five datasets to assess the impacts arising from the dataset, evaluation approaches, and the inherent robustness of models in various program generation contexts. 
This approach allows us to provide deeper insights into how these factors collectively influence the reliability and performance of language models in generating code.

\section{Conclusion}
\label{sec:conlcusion}

In this work, we investigated the reliability and explainability of language models to perform automated program generation.
Our study underlined the following practical conclusion:
\begin{itemize}
\item The performances and key findings of prior works in automated program generation can be duplicated.
\item Automated program generation models are not reliable. Prior works suffer from serious reliability concerns, resulting in unrealistic performance evaluation;
\item Explainable AI approaches are a promising approach to help us understand program generation models.
We discover that program generation models capture distinct patterns but dentifiers and keywords are consistently assigned higher importance scores.
\end{itemize}
Further, our study highlighted two meaningful insights.
Firstly, our study demonstrates that prior program generation approaches do not have perfect robustness and reliability, pointing to opportunities for future improvement.
On the other hand, explainable AI approaches help us better understand model inference behaviors.

\section*{acknowledgment} 
Chakkrit Tantithamthavorn was supported by the Australian Research Council's Discovery Early Career Researcher Award (DECRA) funding scheme (DE200100941). 

\bibliographystyle{ACM-Reference-Format}
\bibliography{sample-base}

\appendix

\end{document}